\RequirePackage{fix-cm}
\documentclass[smallcondensed]{svjour3}   
\smartqed  
\usepackage{graphicx}
\usepackage{amssymb}
\usepackage{pifont}
\usepackage{amsmath}
\usepackage{multirow}
\usepackage{minibox}
\usepackage{enumerate}
\usepackage{mathtools}
\usepackage{wasysym}

\journalname{Nanoscale Research Letters}
\begin{document}

\def\crta{\vrule height1.41ex depth-1.27ex width0.34em}
\def\dj{d\kern-0.36em\crta}
\def\Crta{\vrule height1ex depth-0.86ex width0.4em}
\def\Dj{D\kern-0.73em\Crta\kern0.33em}
\dimen0=\hsize \dimen1=\hsize \advance\dimen1 by 40pt

\title{Can Two-Way Direct Communication
Protocols Be\\ Considered Secure?}

\titlerunning{Can Two-Way QKD Be Considered Secure?}

\author{Mladen Pavi{\v c}i{\'c}}

\authorrunning{Pavi\v ci\'c} 

\institute{M. Pavi{\v c}i{\'c} \at
Center of Excellence for Advanced Materials (CEMS),  
Ru{\dj}er Bo\v skovi\'c Institute, Research Unit Photonics and Quantum
Optics, Zagreb, Croatia and Nanooptics, Department of Physics,
Humboldt-Universit{\"a}t zu Berlin, Germany. \email{mpavicic@irb.hr}}

\date{Received: Jul 6, 2017 / Accepted: Sep 11, 2017}

\maketitle

\begin{abstract}
We consider attacks on two-way quantum key distribution protocols in
which an undetectable eavesdropper copies all messages in the message
mode. We show that under the attacks there is no disturbance in the
message mode and that the mutual information between the sender and
the receiver is always constant and equal to one. It follows that
recent proofs of security for two-way protocols cannot be considered
complete since they do not cover the considered attacks.  
\end{abstract}

\section*{Introduction}
\label{intro}
Quantum cryptography, in particular quantum key distribution 
(QKD) protocols, offers us, in contrast to the classical one,
provably unbreakable communication based on the 
quantum physical properties of the information carriers 
\cite{elliot-DARPA,sasaki-tokyo-qkd-10,peev-zeilinger-njp09}.
So far, the implementations were mostly based on the BB84 
protocol~\cite{bb84} which is unconditionally secure provided the 
quantum bit error rate (QBER) is low enough. However, 
QBER in BB84-like protocols might be high and since we cannot 
discriminate eavesdropper's (Eve's) bit flips from bit flips 
caused by losses and imperfections the request of having QBER low 
enough for processing the bits is often difficult to satisfy.  
E.g., 4-state BB84 with more than 11\% \cite{scarani-09} 
and 6-state BB84 \cite{bruss} with more than 12.6\% \cite{scarani-09}
of disturbance ($D$) have to be aborted ($D$ is defined as the
percentage of polarization-flips caused by Eve, maximum being 0.5).
Since $D$ cannot be discriminated from the inherent QBER in the line,
these levels of total QBER are insecure (mutual information between
the sender (Alice) and Eve ($I_{AE}$) surpasses the one between Alice
and the receiver (Bob) ($I_{AB}$): $I_{AE}>I_{AB}$ for $D>0.11, 0.126$,
respectively) and therefore cannot be carried out just because Eve
{\em might\/} be in the line. 

In search for more efficient protocols, two-way protocols were proposed
and implemented. In particular, entangled photon two-way protocols
based on two \cite{bostrom-felbinger-02} (also called a
{\em ping-pong\/} (pp) protocol) and four ($\Psi^\mp,\Phi^\mp$)
\cite{cai-li-04} Bell states, on the one hand and a single photon
deterministic Lucamarini-Mancini (LM05) protocol, on the other
\cite{lucamarini-05,lucamarini-mancini-13}. Several varieties, 
modifications, and generalisations of the latter protocol are given
in \cite{henao-15,khir-12,shaari-mancini-15,pirandola-nat-08}.
Two varieties were implemented in \cite{cere-06} and 
\cite{kumar-lucamarini-08}. The former pp protocol was implemented
by Ostermeyer and Walenta in 2008~\cite{ostermeyer-08} while the 
protocol with four Bell states cannot be implemented with linear
optics elements \cite{luetkenhaus-99,vaidman99}. In the
aforementioned references various security estimations have been
obtained. 

In \cite{lu-cai-11} Lu, Fung, Ma, and Cai provide a security proof
of an LM05 deterministic QKD for the kind of attack proposed in
\cite{lucamarini-05,lucamarini-mancini-13}. Nevertheless,
they claim it to be a proof of the unconditional security of LM05.
In \cite{han-14} Han, Yin, Li, Chen, Wang, Guo, and Han provide
a security proof for a modified pp protocol and prove its security
against collective attacks in noisy and lossy channel.

All considerations of the security of two-way protocols 
assume that Eve attacks each signal twice, once on the way from
Bob to Alice, and later on its way back from 
Alice to Bob, and that, in doing so, she disturbs the signal in the
message mode. However, as we show below, there are other attacks in
which an undetectable Eve encodes Bob's signals according to
Alice's encoding of a decoy signal sent to her and later on read by
Eve. 

In this paper we show that in the two-way deterministic QKD protocols
under a particular intercept and resend attack an undetectable Eve
can acquire all messages in the message mode (MM) and that the
mutual information between Alice and Bob is constant and equal to one.
That means that the security of the protocols cannot be established
via standard procedures of evaluating the secret fraction of key
lengths.

\section*{Methods}
\label{sec:methods}

We analyze the attacks on two different two-way QKD protocols:
entangled photon and single photon ones. In particular, we
elaborate on the procedure which enables Eve to read off all
the messages in the message mode while remaining undetectable. 
Subsequently, we carry on a security analysis, so as to calculate
mutual information between Alice and Eve, as well as between
Alice and Bob, as a function of the disturbance that Eve
might introduce while eavesdropping.
Eventually, we apply the obtained results on the procedure
which aims at proving an unconditional security of two-way
protocols.

\section*{Results and Discussion}
\label{sec:results}

\subsection*{Entangled Photon Two-Way Protocols}
\label{subsec:entangled}

We consider an entangled-photon two-way protocol based on two Bell
states (pp protocol) \cite{bostrom-felbinger-02}.
Bob prepares entangled photons in one of the Bell
states  and sends one of the photons to Alice
and keeps the other one in a quantum memory. Alice either returns the
photon as is or acts on it so as to put both photons into another Bell
state. The Bell states she sends in this way are her messages to Bob.
Bob combines the photon he receives from Alice with the one he kept
and at a beam splitter (BS) he decodes Alice's messages. Such messages
are said to be sent in a {\em message mode\/} (MM). There is also a
control mode (CM) in which Alice measures Bob's photon. She announces
switching between the modes over a public channel as well as the
outcomes of her measurements in CM. 

We define the Bell basis as a basis consisting of two Bell
states
\begin{eqnarray}
|\Psi^\mp\rangle=\frac{1}{\sqrt{2}}(|H\rangle_1|V
\rangle_2\mp|V\rangle_1|H\rangle_2),
\label{eq:bell-states} 
\end{eqnarray}
where $|H\rangle_i$ ($|V\rangle_i$), $i=1,2$, represent horizontal
(vertical) polarized photon states.

\begin{figure}[ht]
  \begin{center}
(a)\includegraphics[width=0.29\textwidth]{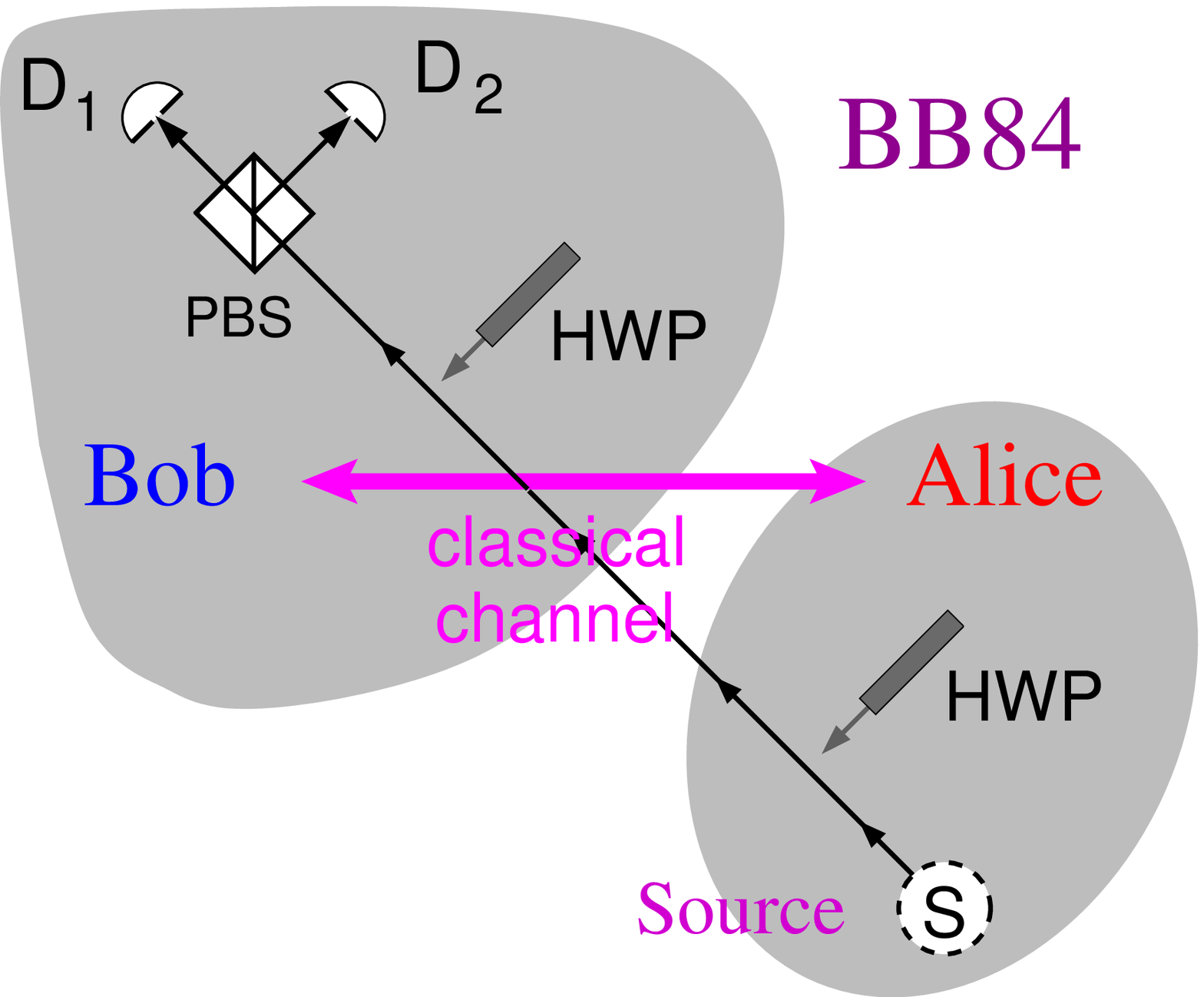}
(b)\includegraphics[width=0.29\textwidth]{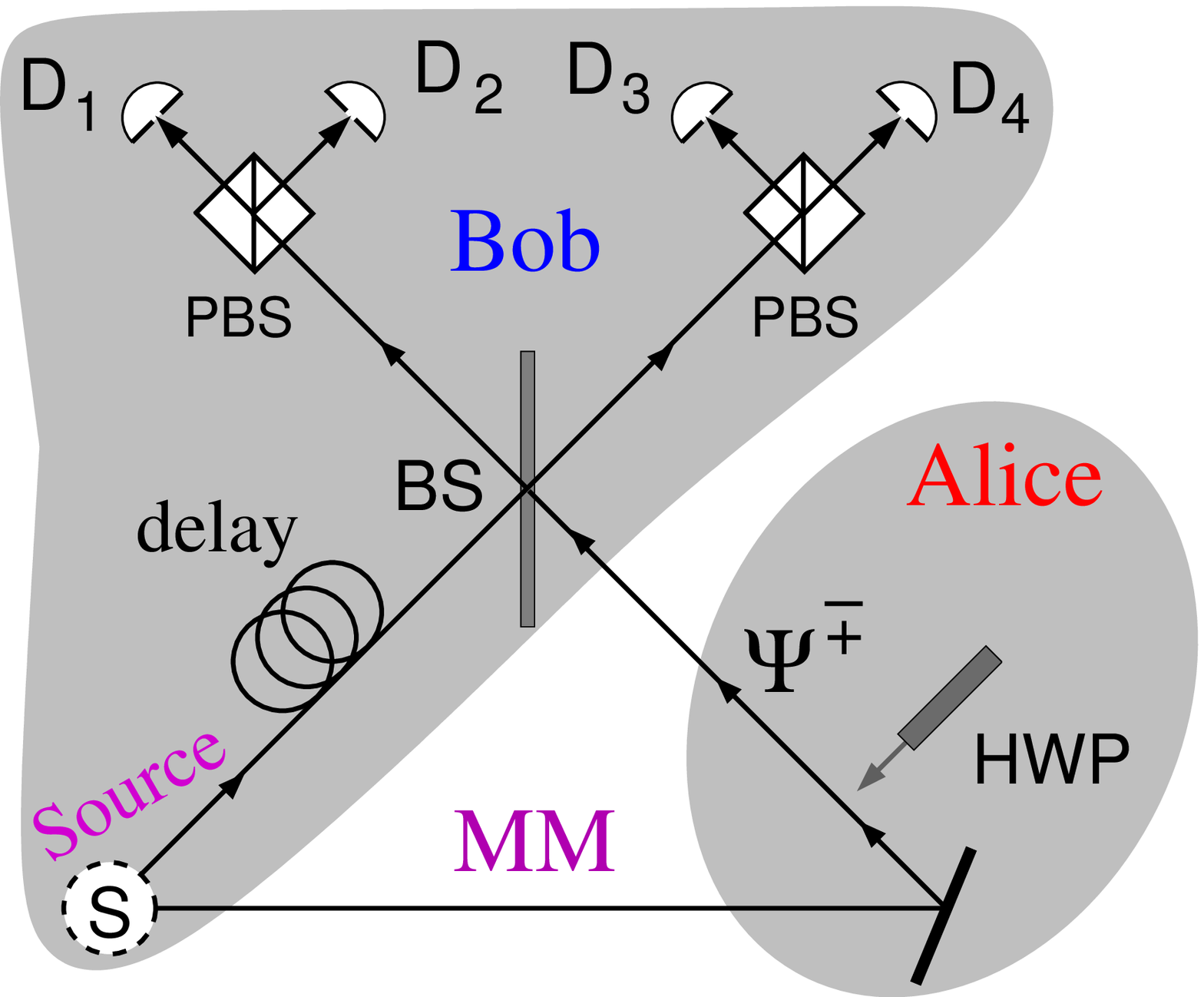}
(c) \includegraphics[width=0.29\textwidth]{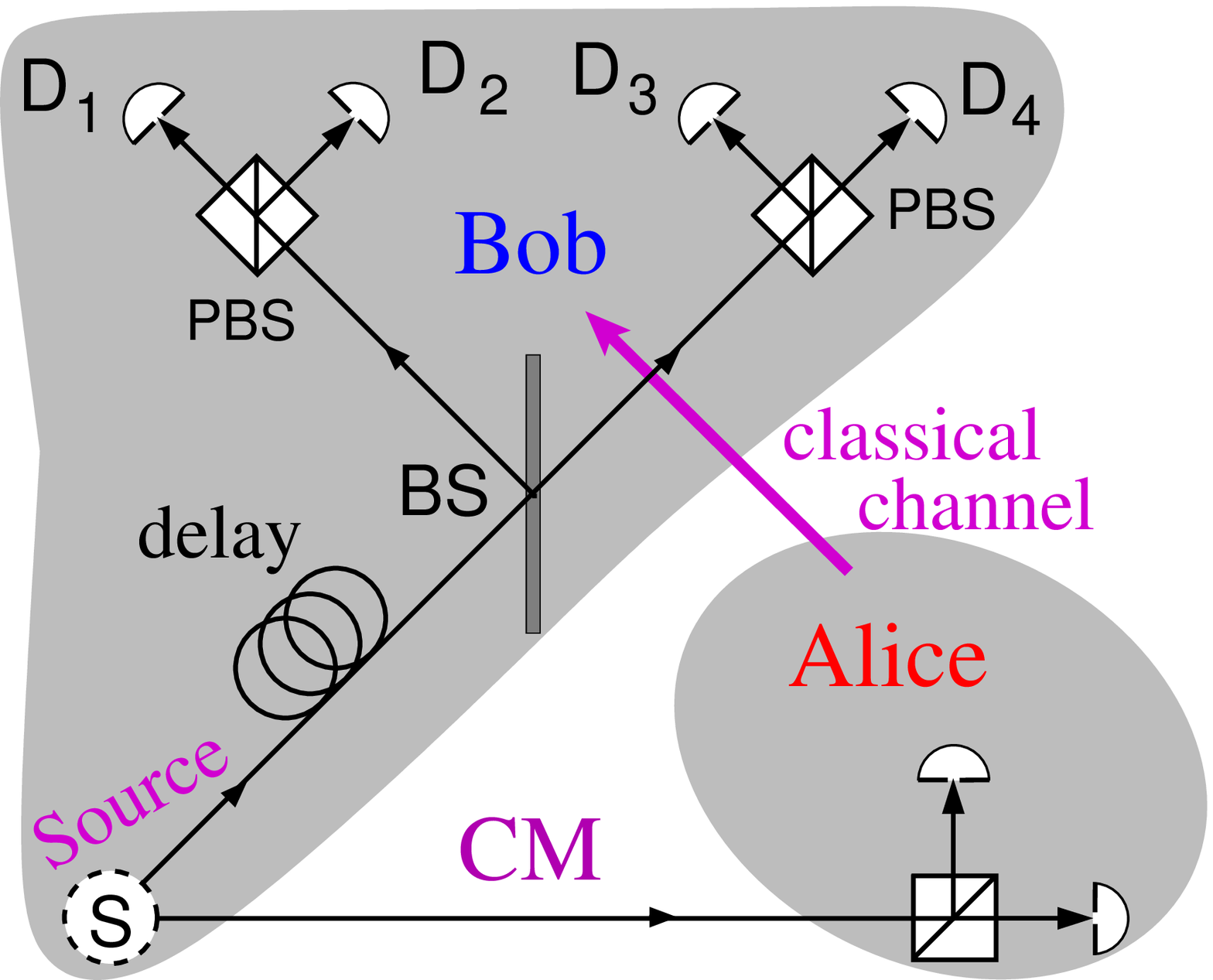}
  \end{center}
  \caption{(a) BB84---classical and quantum channels are merged;
    (b) MM of the pp protocol---quantum channel;
    (c) CM of the pp protocol---classical channel.}
  \label{fig:bell-setup}
  \end{figure}

Photon pairs in the state $|\Psi^-\rangle$ are generated by a
down-converted entangled photon source. To send $|\Psi^-\rangle$
state Alice just returns her photon to Bob. To send $|\Psi^+\rangle$
she puts a half-wave plate (${\rm HWP}(0^\circ)$) in the path of her
photon, as shown in Fig.~\ref{fig:bell-setup}(b). The HWP changes the
sign of the vertical polarization. 

At Bob's BS the photons in state $|\Psi^-\rangle$ will split and
those in state $|\Psi^+\rangle$ will bunch together. 

\begin{figure}[ht]
  \begin{center}
    \includegraphics[width=0.9\textwidth]{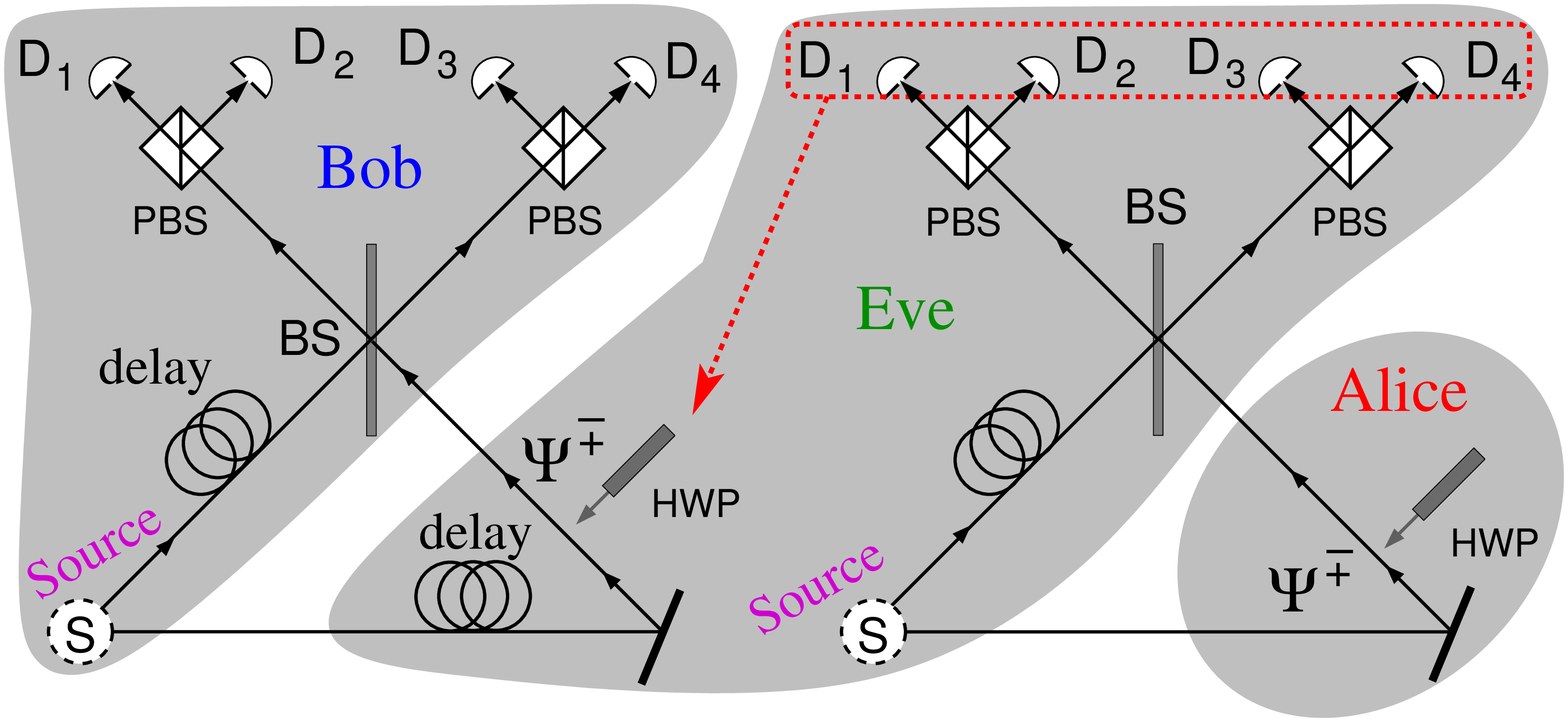}
  \end{center}
  \caption{Nguyen's attack \cite{nguyen-04} by which Eve is able to
    deterministically copy every one of the Bell-state messages in the
    pp protocol \cite{bostrom-felbinger-02}.}
  \label{fig:bell-attack}
  \end{figure}

Eve carries out her attack, designed by Nguyen \cite{nguyen-04}, as
follows. She first puts Bob's photon in a quantum memory and make use
of a copy of Bob's device to send Alice a photon from a down-converted
pair in state $|\Psi^-\rangle$ as shown
in Fig.~\ref{fig:bell-attack}. When Eve receives the photon from Alice
she combines it with the other photon from the pair and determines the
Bell state in the same way Bob would. She uses this result to generate
the same Bell state for Bob by putting the appropriate HWPs in
the path of Bob's photon. 

Thus, Eve is able to copy every single message in MM and therefore
sending of messages in MM is equivalent to sending of plain text
``secured'' by CM. We will come back to this point later on.

Here we stress that photons cover four times the distance they cover
in BB84. So, if the probability of a photon to be detected over only
Bob-Alice distance is $p$, the probability of being detected over
Bob-Alice-Bob distance will be $p^4$ which with the exponentially
increasing losses over distance also exponentially decreases the
probability of detecting the disturbance Eve introduces in CM.

\subsection*{Single Photon Two-Way Protocols}
\label{subsec:single}

\begin{figure}[ht]
\center
(a) \includegraphics[width=0.3\textwidth]{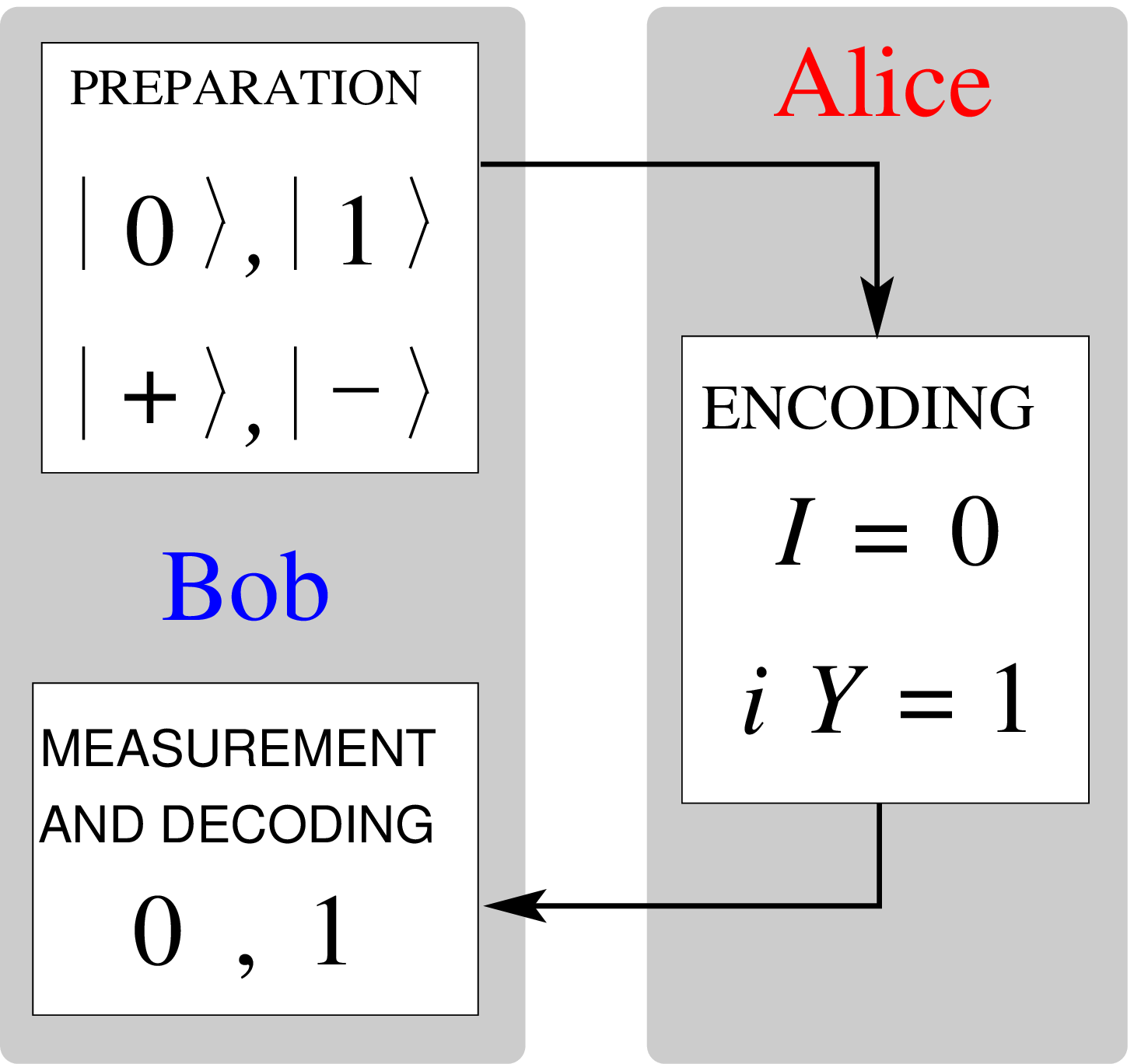}\qquad
(b) \includegraphics[width=0.3\textwidth]{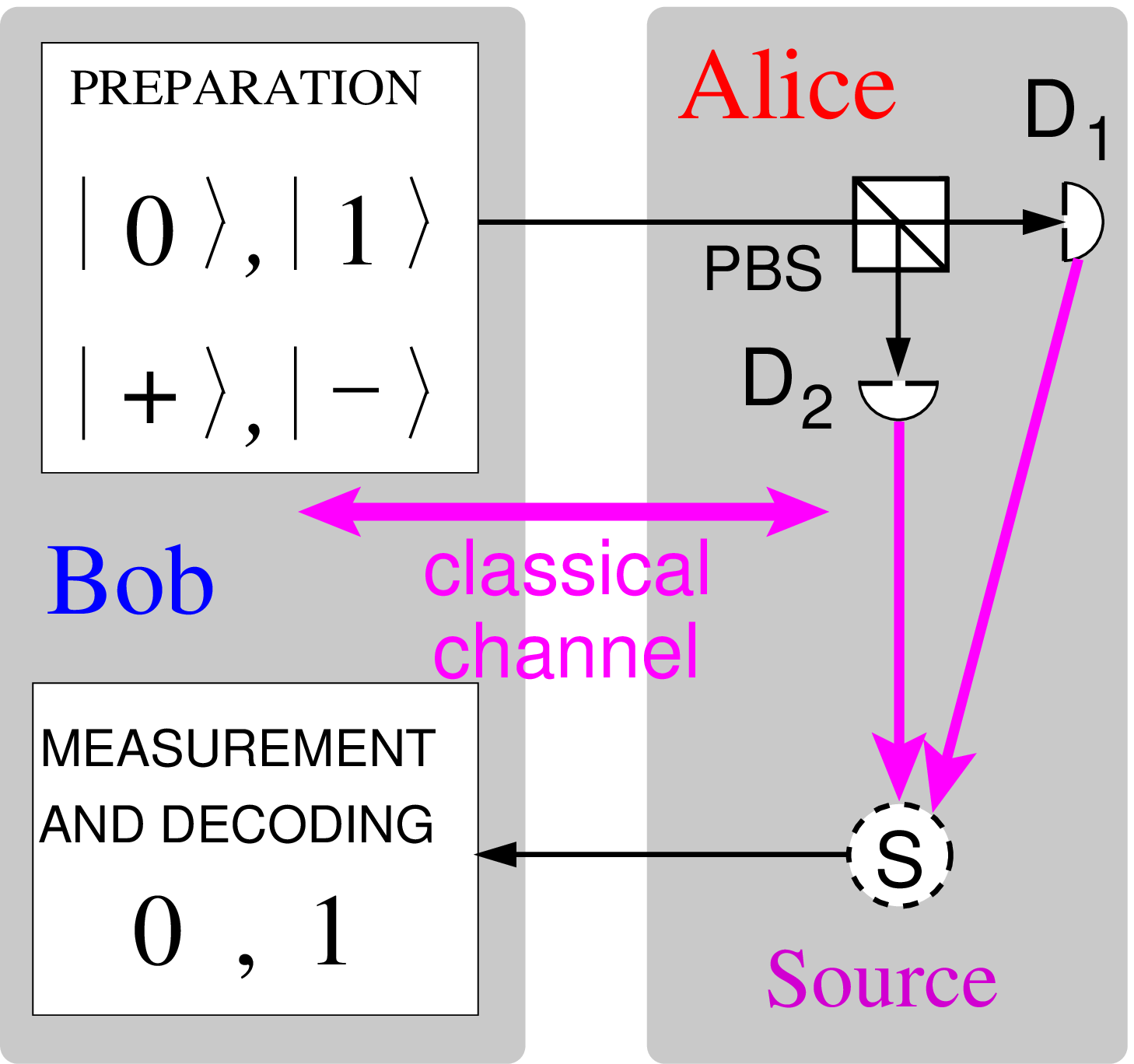}
\caption{(a) MM of LM05 protocol according to
  \cite[FIG.~1]{lucamarini-05}; (b) CM of LM05.} 
\label{fig:lm05}
\end{figure}

We start with a brief presentation of the LM05 protocol
\cite{lucamarini-phd-03,lucamarini-05}. As shown in
Fig.~\ref{fig:lm05}, Bob prepares a qubit in one of the four states
$|0\rangle$,  $|1\rangle$ (the Pauli $\boldsymbol Z$ eigenstates),
$|+\rangle$, or $|-\rangle$ (Pauli $\boldsymbol X$ eigenstates) and
sends it to his counterpart Alice. In the MM she modifies the qubit
state by applying either $\boldsymbol I$, which leaves the qubit
unchanged and encodes the logical {\bf 0}, or by applying
$i{\boldsymbol Y}={\boldsymbol Z}{\boldsymbol X}$, which flips 
the qubit state and encodes the logical {\bf 1}.
($i{\boldsymbol Y}|0\rangle=-|1\rangle$, 
$i{\boldsymbol Y}|1\rangle=|0\rangle$, 
$i{\boldsymbol Y}|+\rangle=|-\rangle$, 
$i{\boldsymbol Y}|-\rangle=-|+\rangle$.) 
Alice now sends the qubit back to Bob who measures it in the same basis
in which he prepared it and deterministically infers Alice’s
operations, i.e., her messages, without basis reconciliation procedure.

\begin{figure}[ht]
\center
\includegraphics[width=0.75\textwidth]{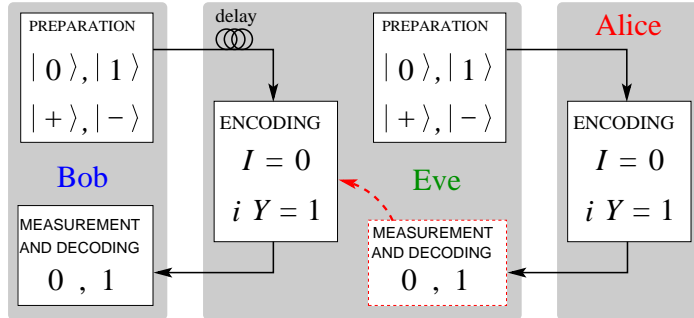}
\caption{Lucamarini's attack on the LM05 protocol
  \cite[p.~61, Fig.~5.5]{lucamarini-phd-03}.}
\label{fig:lm05a}
\end{figure}

The attack on LM05 protocol we consider is proposed by Lucamarini
in \cite[p.~61, Fig.~5.5]{lucamarini-phd-03}. It is shown in
Fig.~\ref{fig:lm05a}. Eve delays Bob's photon (qubit) in a fiber
spool (a quantum memory) and sends her own decoy photon in one of
the four states $|0\rangle$, $|1\rangle$, $|+\rangle$, or
$|-\rangle$ to Alice, instead. Alice encodes her message via
$\boldsymbol I$ or $i{\boldsymbol Y}$ and sends the photon back.
Eve measures it in the same basis in which she prepared it, reads
off the message, encodes Bob's delayed photon via ${\boldsymbol I}$,
if she read {\bf 0}, or via $i{\boldsymbol Y}$, if she read {\bf 1},
and sends it back to Bob. 

Eve never learns the states in which  Bob sent his photons but that 
is irrelevant in the MM since only polarization flipping or not 
flipping encode messages. Alice also need not know Bob's
states \cite{lucamarini-05}. This means that, Eve could only be
revealed in CM in which Alice carries out a projective measurement
of the qubit along a basis randomly chosen between ${\boldsymbol Z}$
and ${\boldsymbol X}$, prepares a new qubit in the same state as
the outcome of the measurement, sends it back to Bob, and reveals
this over a classical public channel
\cite{lucamarini-05}, as shown in Fig.~\ref{fig:lm05}

Here, it should be stressed that photons in LM05 cover twice the
distance they cover in BB84. So, if the probability of a photon to
be detected over only Bob-Alice distance is $p$, the probability of
being detected over Bob-Alice-Bob distance will be $p^2$ and Eve
would be able to hide herself in CM exponentially better than in BB84.

\subsection*{Security of Two-Way Protocols}
\label{subsec:security}

In a BB84 protocol with more 11\%\ of disturbance the mutual information
between Alice and Eve $I_{AE}$ is higher than the mutual information
between Alice and Bob $I_{AB}$ and one has to abort it. 

For our attacks, there is no disturbance ($D$) that Eve induces in MM 
and the mutual information between Alice and Bob is equal to unity. 
\begin{eqnarray}
I_{AB}=1.
\label{eq:m-i}  
\end{eqnarray}
Therefore, unlike in BB84, $I_{AB}$ and $I_{AE}$ are 
not functions of $D$ and that prevents us from proving the 
security using the standard approach.  

Also, in a realistic implementation there is no significant $D$ in MM,
either. When Bob, e.g., sends a photon in $|H\rangle$ state and Alice
does not change it, then Bob will detect $|H\rangle$ with a probability
close to 1, with or without Eve, and independently of distance. The only
QBER which depends on the fiber length is the one that stems from the
dark counts of detectors \cite{stucki-gisin-02}. In a recent
implementation of a one-way QKD the total QBER was under 2\%\ over a
250 km distance \cite{korzh-gisin15}. We can practically completely
eliminate the dark counts, and therefore any uncontrolled polarization
flips, by making use of superconducting transition edge sensor (TES)
photon detectors. The highest efficiency of such detectors is currently
over 98\%~\cite{lita-nam-10,nam-apl13,fukuda-11} and their dark count
probability is practically zero.

For BB84, and practically all one-way one-photon protocols recently
implemented or considered for implementation, the security of the
protocols are evaluated via the critical QBER by calculating the
secret fraction \cite{scarani-09}
\begin{eqnarray}
r=\lim_{N\to\infty}\frac{l}{n}=I_{AB} - I_{AE}
\label{eq:s-f}  
\end{eqnarray}
where $l$ is the length of the list making the final key and $n$ 
is the length of the list making the raw key, 
$I_{AB}=1+D\log_2D+(1-D)\log_2(1-D)$ and
$I_{AE}=-D\log_2D-(1-D)\log_2(1-D)$ and their intersection yields
$D=0.11$. Equivalently, $r=1+2D\log_2D+2(1-D)\log_2(1-D)$ goes down
to 0 when $D$ reaches 0.11.

We do not have such an option for our attacks on two-way
protocols since it follows from Eqs.~(\ref{eq:m-i}) and
(\ref{eq:s-f}) that $r$ is never negative. Actually it approaches
0 only when Eve is in the line all the time.

\begin{figure}[ht]
  \begin{center}
    \includegraphics[width=0.49\textwidth]{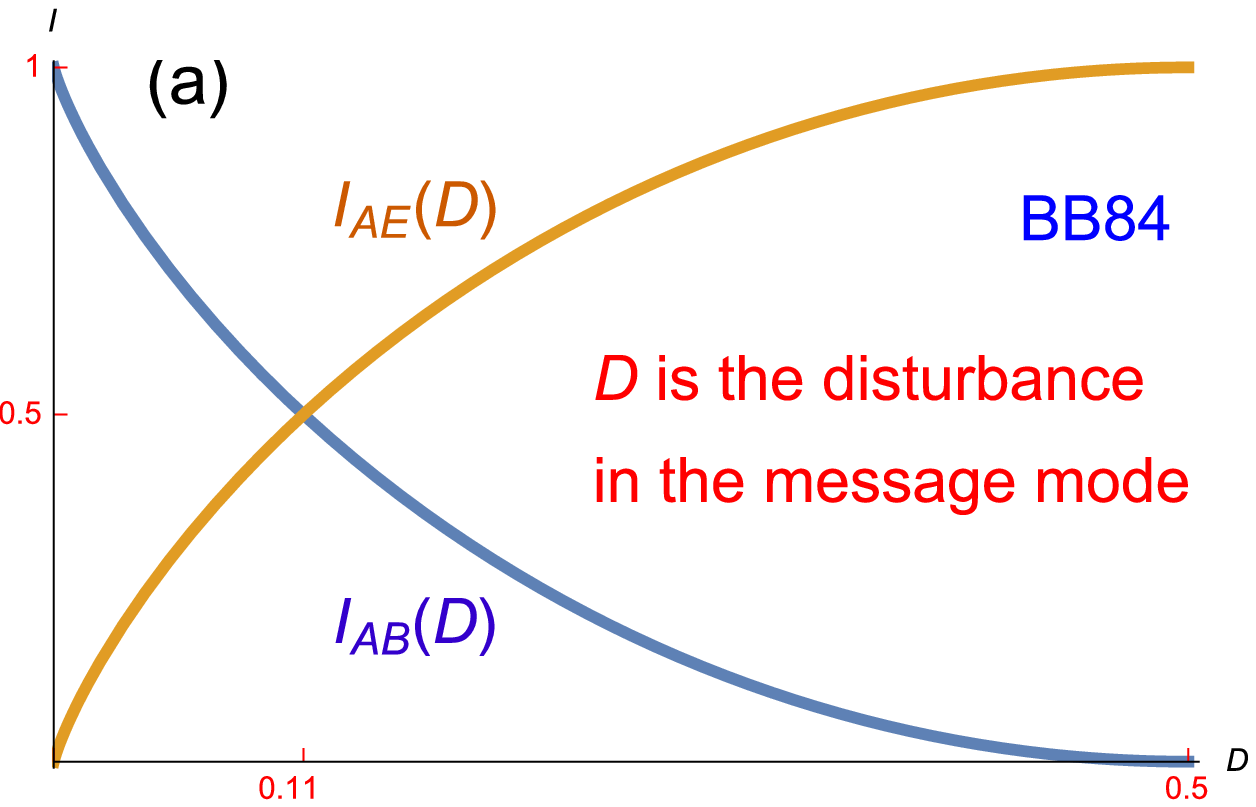}
        \includegraphics[width=0.49\textwidth]{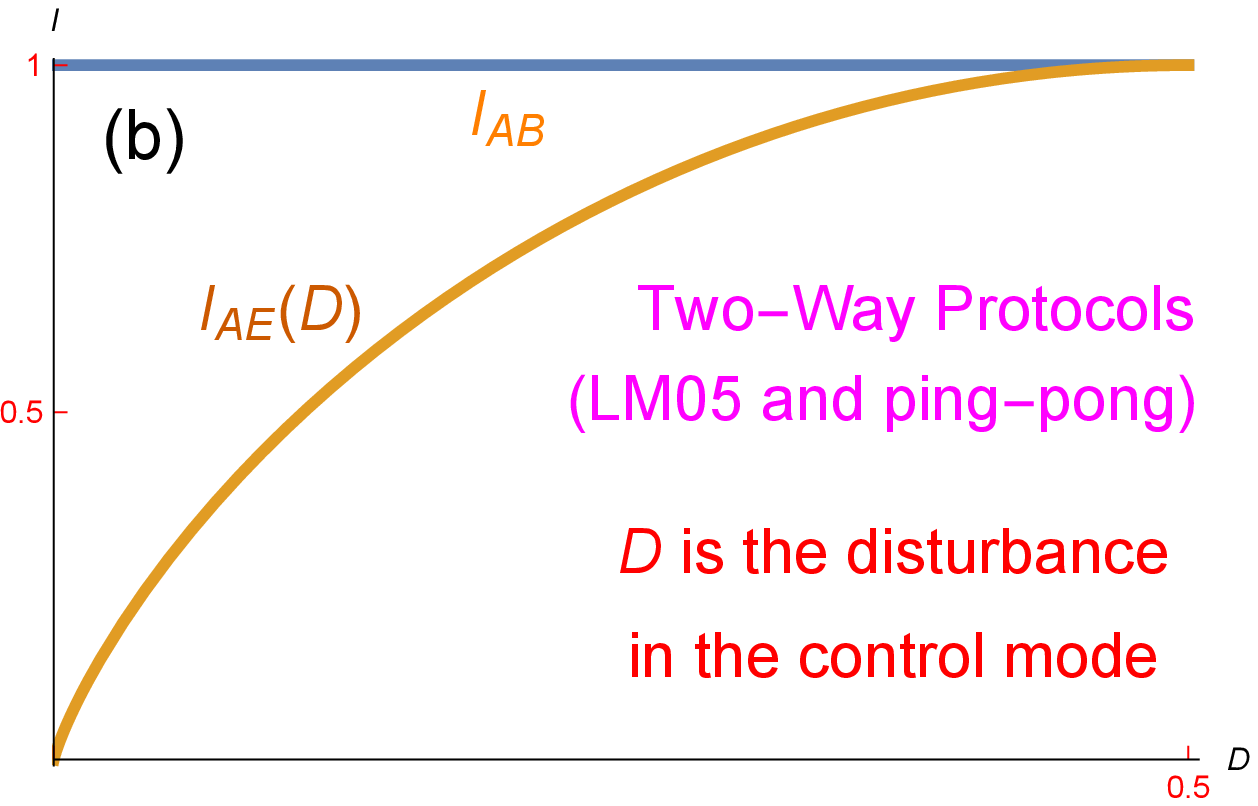}
  \end{center}
  \caption{Mutual information plots for (a) the one-way probabilistic
    protocol BB84 vs.~(b) two-way deterministic protocols with either
    entangled Bell states or with LM05-like single photon states.
    Essential difference between them is that in (a) Eve causes
    polarization flips in the message mode, while in (b) Eve ideally
    does not cause any flip in the message mode.}
  \label{fig:dist}
  \end{figure}

Since $D$ is not related to MM mode in any way it is on Alice
and Bob to decide after which $D$ they would abort their transmission.
However, whichever $0\le D\le 0.5$ they choose
$I_{AB}-I_{AE}$ shall always be non-negative and they will
not have a {\em critical\/} $D$ as in BB84 where the curves $I_{AB}(D)$
and $I_{AE}(D)$ intersect for $D=0.11$ in MM as shown in
Fig.~\ref{fig:dist}(a). For two-way deterministic protocols, the level
of $D$, which is defined in CM (and not in MM), has no effect on
$I_{AB}$, i.e., there is no difference whether $D=0$ or $D=0.5$, as
shown in Fig.~\ref{fig:dist}(b); $0\le D < 0.5$ would only mean that
Eve is not in the line all the time, but Bob always gets full
information from Alice: when Eve is not in the line, because she not
in the line, and when Eve is in the line, because she faithfully
passes all Alice's messages to Bob. 

We can assume that Eve snatches only a portion of messages so as to
keep QBER in CM at a low level (and have $I_{AE}\le 1$) which would be
acceptable to Alice and Bob. With that in mind, we can try to carry
out the security evaluation for our attack and verify whether the
proofs of unconditional security carried out for other kind of attack
on LM05 in \cite{lu-cai-11,lucamarini-mancini-13} might apply to it as 
well. 

In the aforementioned security proof \cite{lu-cai-11}, which is
claimed to be {\em unconditional}, the authors first, in Sec.~III.A,
claim that Eve has to attack the qubits in both the
Bob-Alice and  Alice-Bob channels in to gain Alice’s key bits and
in Sec.~III.B, Eq.~(1,3) they assume that Eve reads off Bob's qubit
and induces a disturbance in the message mode in both Bob-Alice and
Alice-Bob channels (error rate {\em e}; last paragraph of Sec.~III.B
and 1st paragraph of Sec.~III.F).  

However, in the considered attacks Eve does not {\em measure\/} Bob's
qubits. She just stores them in a quantum memory. She sends her own
qubits to Alice and reads off whether she changed them ($Y$) or not
($I$). Then she applies $Y$ or $I$ to stored Bob's qubits and sends
them back to him. Consequently she does not induce any disturbance in
the Alice-Bob channel, either. Also she does not make use of any
ancillas as in \cite{lu-cai-11}. Therefore, the analysis of getting
the key bits carried out in \cite{lu-cai-11} is inapplicable to our
attack. 

Hence, since the proof of security presented in \cite{lu-cai-11}
applies only to the attack considered in it and not to the above
Lucamarini's attack, it is not universal, i.e., it cannot be
considered {\em unconditional\/}.

Let us now consider whether some standard known procedure can be 
used to establish the security of LM05 protocol. In the protocol,
we have neither sifting nor any error rate in the message mode.   
So, the standard error reconciliation cannot be applied either. 

The only procedure we are left with to establish the security is 
the privacy amplification. When Eve possesses just a fraction of data 
she will loose trace of her bits and Alice and Bob's ones will shrink. 
Eve might be able to recover data by guessing the bits she misses and 
reintroduces all bits again in the hash function. If unsuccessful her 
information will be partly wiped away. However, Alice and Bob meet 
a crucial problems with designing their security procedure (e.g., 
hash function) which would guarantee that Eve is left with no 
information about the final key. They do not have a critical amount 
of Eve's bits as in BB84 (11\%) which are explicitly included in 
the equations of the privacy amplification procedure 
\cite{bennett-uncond-sec-ieee-95}.  

In a word, the privacy which should be {\em amplified\/} is not 
well defined. To design a protocol for such a ``blind'' privacy 
amplification is a complex undertaking 
\cite{bennett-uncond-sec-ieee-95} and it is a question whether
sending of---in effect---plain text via MM secured by occasional 
verification of photon states in CM offers us any advantage over 
or a better security than the BB84 protocol. 

In Table \ref{T:comp} we list the properties of a BB84-like protocol 
under an arbitrary attack vs.~two-way protocols under the above 
attacks, which seem to indicate that it would be hard to answer
the aforementioned question in the positive.

\begin{table}[ht]
\center
\setlength{\tabcolsep}{3pt}
\begin{tabular}{|c|c|c|c|}
\hline
&BB84&pp&LM05
\\
\hline 
type
&probabilistic
&deterministic
&deterministic
\\
mode(s)
&message (MM)
&\minibox{message (MM)\\ + control (CM)}
&\minibox{message (MM)\\ + control (CM)}
\\
security
&QBER of MM
&QBER of CM
&QBER of CM
\\
secure
&for QBER $<$ 11\%
&no/unknown
&no/unknown
\\
disturbance
&$0\le D\le 0.5$ in MM
&\minibox{$D=0$ in MM,\\ $0\le D\le 0.5$ in CM}
&\minibox{$D=0$ in MM,\\ $0\le D\le 0.5$ in CM}
\\
\minibox{critical\\ disturbance}
&$D=0.11$
&\minibox{indeterminable ---\\ dependent on inherent\\ QBER of the system}
&\minibox{indeterminable ---\\ dependent on inherent\\ QBER of the system}
\\
\minibox{mutual\\ information}
&\minibox{$I_{AB}=1+D\log_2D$\\ $+(1-D)\log_2(1-D),$\\ 
$I_{AE}=-D\log_2D$\\ $-(1-D)\log_2(1-D)$}
&\minibox{$I_{AB}=1$,\\ $0\le I_{AE}\le 1$}
&\minibox{$I_{AB}=1$,\\ $0\le I_{AE}\le 1$}
\\
\minibox{photon\\ distance}
&$L$
&4$L$
&2$L$
\\
\minibox{trans-\\ mittance}
&$\cal T$
&${\cal T}^4$
&${\cal T}^2$
\\
\hline
\end{tabular}
\caption{Properties of an BB84-like protocol under an arbitrary attack
  compared with properties of pp-like and LM05-like protocols under the
  attack presented in the paper. $0\le I_{AE}\le 1$ simply means that
  Eve might decide not to be in the line only a fraction of time. If
  she was in the line all the time, we would have $I_{AE}=1$.} 
\label{T:comp}
\end{table}

\section*{Conclusion}
\label{sec:conclusion}

To summarise, we considered deterministic attacks on two kinds of
two-way QKD protocols (pp with entangled photons and LM05
with single photons) in which an undetectable Eve can decode all
the messages in the message mode (MM) and showed that the mutual 
information between Alice and Bob is not a function of disturbance
but is equal to unity no matter whether Eve is in the line or not.
Eve induces a disturbance ($D$) only in the control mode (CM) and
therefore the standard approach and protocols for estimating and
calculating the security are not available since they all assume
the presence of $D$ in MM. As a result, a critical $D$ cannot be
determined, the standard error correction procedure cannot be
applied for elimination of Eve's information, the efficiency
of the privacy amplification is curtailed, and the unconditional
security cannot be considered proved. In a way, Alice's sending of
the key is equivalent to sending an unencrypted plain text
``secured'' by an unreliable indicator of Eve's presence and such
protocols cannot be considered for implementation at least not before
one proves or disproves that a novel kind of security procedures for
such deterministic attacks can be designed.

We stress that for deciding whether a protocol is unconditionally
secure or not it is irrelevant whether Eve can carry out attacks
which are more efficient than the attacks considered above, for a
chosen $D$ in $CM$. A proof of unconditional security should cover
them all. 

\begin{acknowledgements}
Financial supports by the Alexander von Humboldt Foundation and
the DFG (SFB787) are acknowledged. Supports by the Croatian Science
Foundation through project IP-2014-09-7515 and by the Ministry of 
Science and Education of Croatia through Center of Excellence CEMS 
are also acknowledged.
\end{acknowledgements}

\bibliographystyle{spmpsci}      

\end{document}